\documentclass[preprint, showpacs,preprintnumbers,amsmath,amssymb,nofootinbib]{revtex4}
\usepackage{mathrsfs}
\usepackage{amsmath}
\usepackage[colorlinks, linkcolor=blue, citecolor=red, urlcolor=red]{hyperref}
\usepackage{amsfonts}
\usepackage{latexsym}
\usepackage{amsfonts}
\usepackage{graphicx}
\usepackage{epsf}
\usepackage{dcolumn}
\usepackage{bm}
\newcommand{\bea}[1]{\begin{eqnarray}\label{#1}}
 \newcommand{\eea}{\end{eqnarray}}

 \def\gsim{ \lower .75ex \hbox{$\sim$} \llap{\raise .27ex \hbox{$>$}} }
 \def\lsim{ \lower .75ex \hbox{$\sim$} \llap{\raise .27ex \hbox{$<$}} }
\def\/{\over}

\begin{document}

\title{\bf Quantum correction to classical gravitational interaction between two polarizable objects}

\author{   Puxun Wu$^{1,2,3}$, Jiawei Hu$^{2}$ and Hongwei Yu$^{1,2,}\footnote{Corresponding author at hwyu@hunnu.edu.cn}$}
\address{ $^1$Department of Physics and Synergetic Innovation Center for Quantum Effects and Applications, Hunan Normal University, Changsha, Hunan 410081, China\\
$^2$Center for Nonlinear Science and Department of Physics, Ningbo
University,  Ningbo, Zhejiang 315211, China\\
$^3$Center for High Energy Physics, Peking University, Beijing 100080, China}
\begin{abstract}
When gravity is quantized, there inevitably exist quantum gravitational vacuum fluctuations which induce quadrupole moments in gravitationally polarizable objects and produce a quantum correction to the classical Newtonian interaction between them. Here, based upon linearized quantum gravity and the leading-order perturbation theory, we study, from a quantum field-theoretic prospect, this quantum correction  between a pair of gravitationally polarizable objects treated as two-level harmonic oscillators. We find that the interaction potential behaves like $r^{-11}$ in the retarded regime and  $r^{-10}$ in the near regime. Our result  agrees with what were recently obtained in  different approaches. Our study seems to indicate that linearized quantum gravity is robust  in dealing with quantum gravitational effects at low energies.

{\bf keywords:} Gravitational interaction; linearized quantum gravity; vacuum fluctuations

\end{abstract}


\maketitle
%\section{Introduction}
Two gravitational wave signals GW150914~\cite{Ligo} and  GW151226~\cite{GW151226} generated by black hole merging systems were detected recently  by the Laser Interferometer Gravitational-Wave Observatory. This confirms  directly  a prediction of Einstein based on his classical theory of general relativity~\cite{Einstein} regarding the existence of  gravitational waves which are spacetime ripples propagating through the universe.  Quantum mechanically, gravitational interaction is presumably mediated by gravitons when gravity is quantized.  However, a full theory of quantum gravity is elusive at the present.  Even though, general relativity as  an effective field theory provides  a framework to probe the low  energy quantum gavity effects. In this respect, it has been found that there exists a quantum correction to the Newtonian  force between two mass monopoles which behaves as $r^{-3}$~\cite{Donoghue}, and this  is obtained by summing one-loop Feynman diagrams with off-shell gravitons.  

Recently, the quantum gravity correction to classical forces  was extended to include the quadrupole-quadrupole interaction  between a pair of polarizable objects from their induced quadrupole moments due to two-graviton exchange~\cite{Ford16}.  This correction is computed by first finding out the normal field modes which keep the cycle in phase in which at the outset one object is polarized and radiates a gravitational field which polarizes the second and induces a quadrupole of which the gravitational field in return polarizes the first, and then summing up the zero-point energy of  all these normal  modes to get the interaction potential between two quadrupoles. This  is in close analogy to that in the computation  of  the Casimir-Polder and van der Waals (vdW) forces between a pair of atoms from their induced dipole moments due to two photon exchange~\cite{Sernelius}. An advantage of  this method is that  the details of quantization of the gravitational metric are not needed.  

  Although a full theory of quantum gravity is  absent, one can still use linearized quantum gravity to find quantum gravitational corrections to classical physics which an ultimate quantum gravity theory must produce at low energies. One such example is the quantum light-cone fluctuations produced by gravitons propagating on a background spacetime~\cite{Yu99, Yu00, Yu09}.  When gravity is quantized, there inevitably exist quantum gravitational fluctuations which induce quadrupoles in gravitationally polarizable  objects, thus giving rise to a quantum correction to the classical interaction between polarizable objects. In this paper, we present in the framework of linearized quantum gravity a different field-theoretic  approach to the computation of the quantum gravity correction to the classical gravitational interaction between a pair of polarizable objects. 
 Our approach, which  is  parallel to that used by Casimir and Polder in studying  the quantum electromagnetic vacuum fluctuation induced electric dipole-dipole interaction between two neutral atoms in a quantum theory of electromagnetism~\cite{Casimir}, is based upon the leading-order perturbation theory.   Let us note that a rather simple perturbative calculation of   both the retarded and instantaneous electromagnetic interaction can also be found in~\cite{Holstein2001}.

%\section{interactions between gravitational field and polarizable objects} \label{secfieldequ}
The  system we consider consists of two  gravitationally polarizable objects  in a bath of fluctuating quantum gravitational fields in vacuum.  We assume that the two  objects labeled as $A$ and $B$ can be treated as  two-level harmonic oscillators. Then, the total Hamiltonian can be give by
\bea{H}
H=H_{F} + H_{A}+H_{B}+H_{AF} +H_{BF}\ ,
\eea
where $H_{F}$ is the Hamiltonian of gravitational fields, the  Hamiltonian of the object $H_{A(B)}$ takes the form
\bea{}
H_{A(B)}=E_{A(B)}^{0} |0_{A(B)}\rangle\langle 0_{A(B)}|+E_{A(B)}^{1} |1_{A(B)}\rangle\langle 1_{A(B)}|\ ,
\eea
and $H_{A(B)F}$   represents the interactions between the objects and gravitational fields 
\bea{HA}
{H}_{A(B)F}=-\frac{1}{2}  {Q}_{A(B)}^{ij}  {E}_{ij}\;.
\eea
Here the Einstein convention is assumed for repeated indices, Latin indices run from 1 to 3,    $ {Q}_{ij}$ is the gravitational vacuum fluctuation induced quadrupole moment of the object and $E_{ij}=R_{0i0j}$ is the gravito-electric tensor with $R_{\mu\nu\alpha\beta}$ being  the Riemann tensor.  $E_{ij}$, which is defined by an analogy between the linearized Einstein field equations and the Maxwell equations~\cite{Campbell},  determines the tidal gravitational acceleration  between two nearby test particles in the classical Newtonian gravity. 
For a flat background spacetime with a linearized perturbation $h_{\mu\nu}$ propagating upon it, the metric can be expanded as  $g_{\mu\nu}=\eta_{\mu\nu}+h_{\mu\nu}$. From the definition of $E_{ij}$ one can obtain 
\bea{Eij}
E_{ij}=\frac{1}{2}\ddot{h}_{ij}\;,
\eea
where a dot denotes a derivative with respect to time $t$.

In the transverse tracefree (TT) gauge, the gravitational field can be quantized as~\cite{Yu99} 
\bea{hij}
h_{ij}({\bf r}, t)=\sum_{{\bf k}, \lambda}[a_{\lambda}(\omega, {\bf r}) e_{ij}({\bf k}, \lambda) f_{{\bf k}}+H. c.]\;,
\eea
where $H.c.$ denotes the Hermitian conjugate, $a_{\lambda}(\omega, {\bf r}) $ is the gravitational field operator,  $\lambda$ labels the polarization states, $e_{ij}({\bf k}, \lambda)$ are polarization tensors, and 
\bea{fk}
 f_{{\bf k}}=\frac{1}{\sqrt{2\omega (2\pi)^{3}}} e^{i ({\bf k} \cdot {\bf r}- \omega t)}
\eea
is the field mode with 
\bea{wk}
\omega=|{\bf k}|=(k_{x}^{2}+ k_{y}^{2} +k_{z}^{2})^{1/2}\ . 
\eea
Here we choose the units in which $32\pi G=1$ and $\hbar=c=1$, where $G$ is the Newton's gravitational constant.
 The vacuum state of gravitational fields  is defined as 
\bea{}
a_{\lambda}(\omega, {\bf r}) | \{0\}\rangle=0\ ,
\eea
and the single- and two-graviton excited states are
\bea{}
a^{\dagger}_{\lambda_{{\alpha}}}(\omega_{\alpha}, {\bf r}_{\alpha}) |\{0\}\rangle&=&|1^{(\alpha)}\rangle\ , \\
\frac{1}{\sqrt{2}}a^{\dagger}_{\lambda_{{\alpha}}}(\omega_{\alpha}, {\bf r}_{\alpha}) a^{\dagger}_{\lambda_{{\beta}}}(\omega_{\beta}, {\bf r}_{\beta}) |\{0\}\rangle&=& |1^{(\alpha)}, 1^{(\beta)}\rangle \ .
\eea

Since the object-gravitational field coupling is linear in the object and the field variables, each object must  interact with the field at least two times and return to its ground state. Thus,  the position-dependent shift of the ground-state energy between  two objects arises from fourth-order perturbations~\cite{Casimir, Buhmann, Safari}  in the leading-order perturbation theory
\bea{ES}
\Delta E_{AB}=&-&{\sum_{\mathrm{I,II,III}}}' \frac{\langle 0| \hat{H}_{AF}+ \hat{H}_{BF}| \mathrm{I} |\langle\mathrm{I} | \hat{H}_{AF}+ \hat{H}_{BF}| \mathrm{II} \rangle}{(E_{\mathrm{I}}-E_{0})(E_{\mathrm{II}}-E_{0})} \nonumber \\  && \qquad \times \frac { \langle\mathrm{II} | \hat{H}_{AF}+ \hat{H}_{BF}| \mathrm{III} \rangle \langle\mathrm{III} | \hat{H}_{AF}+ \hat{H}_{BF}| 0\rangle}{(E_{\mathrm{III}}-E_{0})}\;,
\eea
where the primed sum means that the ground state of the whole system $|0\rangle= |0_{A}\rangle |0_{B}\rangle|\{0\}\rangle$ is omitted and the summation includes position and frequency integrals. During each interaction, a graviton may either be emitted or absorbed by an object. The intermediate states  $| \mathrm{I} \rangle$ and  $| \mathrm{III} \rangle$ must consist  of  an excited object and a single graviton. While for the intermediate state  $| \mathrm{II} \rangle$ there are three possibilities, i.e.,  both objects in the ground state with two gravitons, both objects excited   with no graviton  or both objects excited   with two gravitons. Therefore, there are ten possible combinations of intermediate states, which are listed in Table.~(\ref{Tab1}).

For the case (1) in Table.~(\ref{Tab1}), substituting $|\mathrm I\rangle$, $|\mathrm{ II}\rangle$ and $|\mathrm{III}\rangle$ into Eq.~(\ref{ES})  yields  
\bea{}
\Delta E_{AB(1)}({\bf r}_{A}, {\bf r}_{B})  
&=& -\frac{1}{16} \int_{0}^{\infty}d\omega \int_{0}^{\infty}d\omega' \bigg(\frac{1}{D_{ \mathrm i}} +\frac{1}{D_{\mathrm {ii}}}   \bigg) \nonumber \\
&& \times \tilde{Q}_{A}^{ij} \ \tilde{Q}_{A}^{*kl}\ \tilde{Q}_{B}^{ab}\ \tilde{Q}_{B}^{*cd}\  G_{ijab}(\omega,  {\bf r}_{A}, {\bf r}_{B}) G_{klcd}(\omega',  {\bf r}_{A}, {\bf r}_{B})\ ,
\eea
where $\tilde{Q}_{A(B)}^{ij}= \langle 0_{A(B)}| {Q}_{A(B)}^{ij} | 1_{A(B)}\rangle$, $\tilde{Q}_{A(B)}^{*ij}= \langle 1_{A(B)}| {Q}_{A(B)}^{ij} | 0_{A(B)}\rangle$, and
\bea{fijkl}
 G_{klcd}(\omega,  {\bf r}_{A}, {\bf r}_{B})=\langle 0| {E}_{kl}(\omega, {\bf r}_{A}) {E}_{cd}(\omega, {\bf r}_{B}) |0\rangle 
 \eea
 is the two-point correlation function of gravito-electric fields.
The expressions of  $D_{\mathrm i}$ and $D_{\mathrm {ii}}$ are given in Tab.~(\ref{Tab1}) in which $\omega_{A(B)}=(\omega_{A(B)}^{1}-\omega_{A(B)}^{0})$ with $\omega_{A(B)}^{1}=E_{A(B)}^{1}$ and $\omega_{A(B)}^{0}=E_{A(B)}^{0}$, which represents the transition frequency  of  the object. The contributions of other cases in Table.~(\ref{Tab1}) to $\Delta E_{AB}$ can be calculated similarly. Summing up all possible intermediate states, we obtain that 
\bea{}
\Delta E_{AB}({\bf r}_{A}, {\bf r}_{B})  
&=& -\frac{1}{16} \int_{0}^{\infty}d\omega \int_{0}^{\infty}d\omega' \sum_{n=\mathrm i}^{\mathrm  {xii} } \frac{1}{D_{n}}  \nonumber \\ && \times \tilde{Q}_{A}^{ij} \ \tilde{Q}_{A}^{*kl}\ \tilde{Q}_{B}^{ab}\ \tilde{Q}_{B}^{*cd}\  G_{ijab}(\omega,  {\bf r}_{A}, {\bf r}_{B}) G_{klcd}(\omega',  {\bf r}_{A}, {\bf r}_{B})\ .
\eea
It is easy to show
\bea{}
\sum_{n=\mathrm i}^{\mathrm { xii}} \frac{1}{D_{n}}= \frac{4(\omega_{A}+\omega_{B}+\omega)}{(\omega_{A}+\omega_{B})(\omega_{A}+\omega)(\omega_{B}+\omega)} \bigg(\frac{1}{\omega+\omega'}-\frac{1}{\omega-\omega'} \bigg)\ .
\eea

Since the  shift of the ground-state energy  is just   the vdW-like potential $U_{AB}({\bf r}_{A}, {\bf r}_{B})$ between two polarizable objects, one has from the above two equations, 
\bea{UAB}
U_{AB}({\bf r}_{A}, {\bf r}_{B})  
&=& -\frac{1}{4} \int_{0}^{\infty}d\omega \int_{0}^{\infty}d\omega' \frac{\omega_{A}+\omega_{B}+\omega}{(\omega_{A}+\omega_{B})(\omega_{A}+\omega)(\omega_{B}+\omega)} \bigg(\frac{1}{\omega+\omega'}-\frac{1}{\omega-\omega'} \bigg) \nonumber \\
&&\qquad \qquad \times \tilde{Q}_{A}^{ij} \ \tilde{Q}_{A}^{*kl}\ \tilde{Q}_{B}^{ab}\ \tilde{Q}_{B}^{*cd}\  G_{ijab}(\omega,  {\bf r}_{A}, {\bf r}_{B}) G_{klcd}(\omega',  {\bf r}_{A}, {\bf r}_{B})\ .
\eea
 Assuming that the objects A and B are isotropically polarizable , we have \bea{Q}
\tilde{Q}_{A(B)}^{ij}  \tilde{Q}_{A(B)}^{*kl} = 2 |\tilde{Q}_{A(B)}^{ij}|^{2} \delta_{ik}\delta_{jl}\equiv \frac{1}{2} \tilde{\alpha}_{A(B)} \delta_{ik}\delta_{jl}\ .
\eea
Here $\tilde{\alpha} (\omega)\equiv \frac{1}{4}|\tilde{Q}_{ij}(\omega) |^{2}$. 
Then, Eq.~(\ref{UAB}) can be simplified as 
\bea{UAB0}
U_{AB}({\bf r}_{A}, {\bf r}_{B}) &=& -\frac{1 }{16(\omega_{A}+\omega_{B})} \int_{0}^{\infty}d\omega \int_{0}^{\infty}d\omega' \;\frac{ \tilde{\alpha}_{A} \tilde{\alpha}_{B}(\omega_{A}+\omega_{B}+\omega)}{(\omega_{A}+\omega)(\omega_{B}+\omega)} \bigg(\frac{1}{\omega+\omega'}-\frac{1}{\omega-\omega'} \bigg) \nonumber \\
&&\qquad \qquad \times  G_{ijab}(\omega,  {\bf r}_{A}, {\bf r}_{B}) G_{ij ab }(\omega',  {\bf r}_{A}, {\bf r}_{B})\ .
\eea
The two-point correlation function $G_{ijab}(\omega,  {\bf r}_{A}, {\bf r}_{B}) $ can be obtained from  $G_{ijab}(  {\bf r}_{A}, {\bf r}_{B},  {t_{A}}, t_{B})$  by  Fourier transform. From Eqs.~(\ref{Eij}), (\ref{hij}), and (\ref{fijkl}), one finds 
\bea{}
 G_{ijkl}({\bf r}, {\bf r}', t, t') &=& \frac{1}{4}\langle 0| \ddot {h}_{ij}( {\bf r}, t) \ddot{h}_{kl}( {\bf r}', t') |0\rangle \nonumber \\
&=&   \frac{1}{4 (2\pi)^{3}}\int d^{3} {\bf k} \sum_{\lambda} e_{ij}({\bf k}, \lambda) e_{kl}({\bf k}, \lambda)  \frac{\omega^{3}}{2} e^{i{\bf k}\cdot ({\bf r-r'})- i\omega(t-t')} \ .\eea
The summation of polarization tensors in the TT gauge gives~\cite{Yu99}
\bea{eijekl}
 \sum_{\lambda}\, e_{ij} ({{\bf k}, \lambda}) e_{kl} ({{\bf k}, \lambda})&=&\delta_{ik}\delta_{jl}
+\delta_{il}\delta_{jk}-\delta_{ij}\delta_{kl}
+\hat k_i\hat k_j \hat k_k\hat k_l+\hat k_i \hat k_j \delta_{kl} \nonumber\\
&&+\hat k_k \hat k_l \delta_{ij}-\hat k_i \hat k_l \delta_{jk}
-\hat k_i \hat k_k \delta_{jl}-\hat k_j \hat k_l \delta_{ik}-\hat k_j \hat k_k \delta_{il}\,,
\eea
where 
\begin{equation}
\hat k_i=\frac{ k_i}{ k}\,.
\end{equation}

Transforming to the spherical coordinate: $\hat{k}_{1}=\sin \theta \cos \varphi$, $\hat{k}_{2}=\sin \theta \sin \varphi$ and $\hat{k}_{3}=\cos \theta $, and letting
\bea{eijekl}
 \sum_{\lambda}\, e_{ij} ({{\bf k}, \lambda}) e_{kl} ({{\bf k}, \lambda})=g_{ijkl}(\theta, \varphi)\,,
\eea
we arrive at
\bea{AG1}
 G_{ijkl}(r, \Delta t)  =   \frac{1}{8(2\pi)^{3} }\int_{0}^{\infty} \omega^{5} d \omega \int_{0}^{\pi} \sin\theta d\theta \int_{0}^{2\pi} d \varphi \  g_{ijkl}(\theta, \varphi)  e^{i \omega (r \cos\theta-  \Delta t)} \ ,\eea
where $r=\Delta {\bf r}= |{\bf r}_{A}- {\bf r}_{B}|$ is the distance between objects A and B.   Fourier transforming  the above expression yields 
\bea{G16}
   G_{ijkl}(\omega,  {\bf r}_{A}, {\bf r}_{B}) & =& \frac{1}{2\pi} \int_{-\infty}^{\infty} d \Delta t e^{i\omega \Delta t} G_{ijkl}( { r}, {\Delta t}) \nonumber \\
  & = &  \frac{\omega^{5}}{8(2\pi)^{3} }  \int_{0}^{\pi} \sin\theta d\theta \int_{0}^{2\pi} d \varphi \  g_{ijkl}(\theta, \varphi)  e^{i w r \cos\theta }\ ,
\eea
 Substituting Eq.~(\ref{G16}) into Eq.~(\ref{UAB0}) and integrating over ($\theta, \varphi, \theta', \varphi'$), we obtain
\bea{UAB2}
U_{AB}(r)  
&=&  -\frac{1 }{2^{9} \pi^{4}(\omega_{A}+\omega_{B}) r^{10}} \int_{0}^{\infty}d\omega \int_{0}^{\infty}d\omega' \;\frac{  \tilde{ \alpha}_{A} \tilde{\alpha}_{B}(\omega_{A}+\omega_{B}+\omega)  }{(\omega_{A}+\omega)(\omega_{B}+\omega)} \bigg(\frac{1}{\omega+\omega'}-\frac{1}{\omega-\omega'} \bigg) \nonumber \\
 &&\quad\quad \times [A_{1}(\omega r, \omega' r)\cos(\omega' r)+B_{1}(\omega r, \omega' r)\sin(\omega' r)]\ ,
\eea
where 
\bea{A1}
A_{1}(x, x') &=& x x' (315 +8 x^{2}x'^{2}-30x^{2}- 30x'^{2})\cos x \\ \nonumber &+& x'(-315+ 135x^2+30 x'^{2} - 18 x^{2} x'^{2}- 3x^{4} + 2 x^{4} x'^{2})\sin x \ ,
\eea
and
\bea{B1}
B_{1}(x, x') &=& x (-315+ 135x'^2+30 x^{2} - 18 x^{2} x'^{2}- 3x'^{4} + 2 x^{2} x'^{4})\cos x  \\ \nonumber &
+&  (315 -135 x^{2} +3 x^{4}- 135 x'^{2}+ 63 x^{2} x'^{2}-3  x^4 x'^2 + 3 x'^4 - 3 x^2 x'^4 +  x^4 x'^4)\sin x \ .
\eea
Since $A_{1}(x, -x')=-A_{1}(x, x')  $ and $B_{1}(x, -x')=B_{1}(x, x') $, Eq.~(\ref{UAB2}) can be re-expressed as 
\bea{UAB3}
U_{AB}(r) &=&   -\frac{ 1 }{2^{9} \pi^{4}(\omega_{A}+\omega_{B}) r^{10}} \int_{0}^{\infty}d\omega\; \frac{ \tilde{ \alpha}_{A} \tilde{\alpha}_{B} (\omega_{A}+\omega_{B}+\omega)  }{(\omega_{A}+\omega)(\omega_{B}+\omega)}  \\   \nonumber
&& \times \int_{-\infty}^{\infty} d\omega' \bigg( \frac{1}{2(\omega+\omega')}+\frac{1}{2(-\omega+\omega')} \bigg) [A_{1}(\omega r, \omega' r)-iB_{1}(\omega r, \omega' r)]e^{i\omega' r} .
\eea
Performing the principle value integral on $\omega'$, one has 
\bea{UAB3}
U_{AB}(r) &=&   -\frac{1}{ 32^{2}\pi^{3} (\omega_{A}+\omega_{B}) r^{10}} \int_{0}^{\infty}d\omega\; \frac{ \tilde{ \alpha}_{A} \tilde{\alpha}_{B}  (\omega_{A}+\omega_{B}+\omega)  }{(\omega_{A}+\omega)(\omega_{B}+\omega)} \\ \nonumber 
&&\times [A_{2}(\omega r) \cos(2\omega r)+B_{2}(\omega r) \sin(2\omega r)]\ ,
\eea
where 
\bea{A2}
A_{2}(x)&=&-630 x+330 x^{3}-42 x^{5}+4x^{7} \ ,\nonumber \\ \nonumber
B_{2}(x)&=&315 -585 x^{2}+ 129 x^{4} - 14 x^{6}+ x^{8} \ .
\eea
Eq.~(\ref{UAB3}) can be re-written as 
\bea{}
U_{AB}(r) &=&   -\frac{1 }{ 32^{2} \pi^{3}  (\omega_{A}+\omega_{B}) r^{10}} \bigg(\int_{0}^{\infty}d\omega\; \frac{   \tilde{\alpha}_{A} \tilde{ \alpha}_{B} (\omega_{A}+\omega_{B}+\omega)  }{(\omega_{A}+\omega)(\omega_{B}+\omega)} \bigg [\frac{A_{2}(\omega r)}{2} +\frac{B_{2}(\omega r)}{2i} \bigg] e^{2i\omega r} \nonumber  \\ 
&&+\int_{0}^{-\infty}d\omega\; \frac{  \tilde{\alpha}_{A} \tilde{\alpha}_{B} (\omega_{A}+\omega_{B}-\omega)  }{(\omega_{A}-\omega)(\omega_{B}-\omega)} \bigg [\frac{A_{2}(\omega r)}{2}  +\frac{B_{2}(\omega r)}{2i} \bigg] e^{2i\omega r} \bigg) \ .
\eea
Simplifying the above equation by contour-integral techniques, we obtain
\bea{UAB31}
U_{AB}(r)  &=&  -\frac{ 1 }{32^{2} \pi^{3}  r^{10}}  \int_{0}^{\infty}d u \  {\alpha}_{A}(iu)  {\alpha}_{B}(iu)[i A_{2}(i u r) + B_{2}(i u r)] e^{-2 u r} \nonumber \\
 &=&  -\frac{ 1 }{32^{2} \pi^{3}  r^{10}}  \int_{0}^{\infty}d u \  {\alpha}_{A}(iu)  {\alpha}_{B}(iu) S(ur) e^{-2 u r} \ .
\eea
where 
\bea{}
{\alpha}_{A(B)}(\omega)=\lim _{\epsilon \rightarrow0^{+}}\frac{ \tilde{\alpha}_{A(B)} \omega_{A(B)}}{\omega_{A(B)}^{2}-\omega^{2}-i \epsilon \omega}=\lim _{\epsilon \rightarrow0^{+}} \frac{1}{4}\frac{ |\tilde{Q}^{ij}_{A(B)}|^{2} \omega_{A(B)}}{\omega_{A(B)}^{2}-\omega^{2}-i \epsilon \omega}\ ,
\eea 
is the object's ground-state polarizability, which is defined in analogy to the electric polarizability of atoms~\cite{Buhmann1} and satisfies 
\bea{}Q_{ij}(\omega)= {\alpha}(\omega) E_{ij} (\omega, \bf r)\ ,\eea and
\bea{}
S(x)=315 + 630 x + 585 x^2  + 330 x^3  + 129 x^4  +  42 x^5 + 14 x^6 + 4 x^7 + x^8 \ .
\eea
 
 In the far regime $r\gg \omega_{A(B)}^{-1}$, since there is an exponential in the integrand in Eq.~(\ref{UAB31}), small values of $u$ provide the dominating contribution. Thus, we can use approximately  the static polarizability $\tilde{\alpha}_{A(B)}(0)$ and obtain
  \bea{FL}
U_{AB}(r) &=&  -\frac{3987}{ 4 \pi (32\pi)^{2}   r^{11}} {\alpha}_{A}(0) {\alpha}_{B}(0) \nonumber \\
&=&  -\frac{3987 \hbar c G^{2}}{ 4\pi r^{11}}  {\alpha}_{A}(0)  {\alpha}_{B}(0)\ .
\eea
 In the near regime  $r\ll \omega_{A(B)}^{-1}$, the integral in Eq.~(\ref{UAB31}) is effectively limited to a region where $e^{-2ur}\simeq 1$ and so all terms in $S(x)$ dependent on  $x$ can be neglected.  Thus, the potential becomes
  \bea{NL}
U_{AB}(r)  &=&  -\frac{ 315}{32^{2} \pi^{3}  r^{10}}   \int_{0}^{\infty}d u \  {\alpha}_{A}(iu)  {\alpha}_{B}(iu) \nonumber   \\
 &=&  -\frac{ 315 \hbar G^{2}}{  \pi   r^{10}}   \int_{0}^{\infty}d u \  {\alpha}_{A}(iu)  {\alpha}_{B}(iu) \ .
\eea
In the second line of Eqs.~(\ref{FL}) and (\ref{NL}), we return to the SI units.  These results  agree with that  obtained with a normal mode evaluation~\cite{Ford16} and  a two-graviton exchange calculation~\cite{Holstein2016} 
 It is interesting to note here that  the qualitative behavior of the effective interaction between two gravitationally polarizable objects can be inferred from a simple dimensional analysis. Dimensionally, the object's polarizability scales as $1/M^5$.  Since the retarded potential energy must go as $M$ and is quadratic in the polarizability, the $r$ dependence must be $ r^{-11}$.  In the instantaneous case there is an additional factor of the excitation energy involved,  so the scaling is as $ r^{-10}$.  

In conclusion,  based upon the quantum theory of linearized gravity and the leading-order perturbation theory, we have studied the quantum gravitational vacuum fluctuation induced quadrupole-quadrupole quantum correction to the classical Newtonian force  between a pair of polarizable objects. We find that the interaction potential decays as $r^{-11}$  in the retarded region and  $r^{-10}$ in the near region.
Our approach parallels that of Casimir and Polder in the investigation of  the quantum electromagnetic vacuum fluctuation induced dipole-dipole interaction between two neutral atoms in a quantum theory of electromagnetism and  suggests that it may be robust to use linearized quantum gravity to study quantum gravitational effects at low energies, such as those associated with quantized gravitational waves in the same way as one usually does with quantized electromagnetic waves.

\acknowledgments  This work was supported by the National Natural Science Foundation of China under Grants No. 11435006, No. 11375092 and No. 11447022;  and the Zhejiang Provincial Natural Science Foundation of China under Grant No. LQ15A050001.

%\begin{widetext}
\begin{table}[ht]
\begin{center}
\begin{tabular}{cllll}
\hline
 Case  & $|\mathrm I\rangle$    &
 \hspace{1ex}
 $|\mathrm{II}\rangle$   &
 \hspace{1ex}
 $\hspace{-1ex}
 |\mathrm{III}\rangle$ &
 Denominator\\
\hline
($1$)
& $|1_A,0_B\rangle |1^{(1)}\rangle$
      &\hspace{1ex} $|0_A,0_B\rangle
        |1^{(2)},1^{(3)}\rangle$
      & \hspace{1ex}$|0_A,1_B\rangle |1^{(4)}\rangle$
      & $D_{\mathrm {i}}=(\omega_A+\omega')
      (\omega'+\omega)(\omega_B+\omega')$,  \\
      {}
      & ${}$
      & ${}$
      & ${}$
      & $D_{\mathrm {ii}}=(\omega_A+\omega')
      (\omega'+\omega)(\omega_B+\omega)$  \\
($2$)
      &$|1_A,0_B\rangle |1^{(1)}\rangle$
      & \hspace{2ex}$|1_A,1_B\rangle |\{0\}\rangle$
      & \hspace{1ex}$|0_A,1_B\rangle |1^{(2)}\rangle$
      & $D_{\mathrm {iii}}=(\omega_A+\omega')
        (\omega_A+\omega_B)
        (\omega_B+\omega)$\\
($3$)
      &$ |1_A,0_B\rangle |1^{(1)}\rangle$
      & \hspace{2ex} $|1_A,1_B\rangle |\{0\}\rangle$
      &\hspace{1ex}$|1_A,0_B\rangle |1^{(2)}\rangle$
      & $D_{\mathrm {iv}}=(\omega_A+\omega')
         (\omega_A+\omega_B)
            (\omega_A+\omega)$\\
($4$)
      & $|1_A,0_B\rangle |1^{(1)}\rangle$
      & \hspace{2ex}$|1_A,1_B\rangle
        |1^{(2)},1^{(3)}\rangle$
      & \hspace{1ex}$|0_A,1_B\rangle |1^{(4)}\rangle$
      & $D_{\mathrm {v}}=(\omega_A+\omega')
         (\omega_A+\omega_B+\omega'+
         \omega)
     (\omega_B+\omega')$\\
($5$)
      & $|1_A,0_B\rangle |1^{(1)}\rangle$
      & \hspace{2ex} $|1_A,1_B\rangle |1^{(2)}, 1^{(3)}\rangle$
      & \hspace{1ex}$|1_A,0_B\rangle |1^{(4)}\rangle$
      & $D_{\mathrm {vi}}=(\omega_A+\omega')
         (\omega_A+\omega_B+
         \omega'+\omega)
     (\omega_A+\omega)$\\
($6$)
      & $|0_A,1_B\rangle |1^{(1)}\rangle$
      & \hspace{2ex}$|0_A,0_B\rangle
        |1^{(2)},1^{(3)}\rangle$
      & \hspace{1ex}$|1_A,0_B\rangle |1^{(4)}\rangle$
            & $D_{\mathrm {vii}}=(\omega_B+\omega')
      (\omega'+\omega)(\omega_A+\omega')$,  \\
{}
      & ${}$
      & ${}$
      & ${}$
           & $D_{\mathrm{viii}}=(\omega_B+\omega')
      (\omega'+\omega)(\omega_A+\omega)$  \\
($7$)
      &  $|0_A,1_B\rangle |1^{(1)}\rangle$
      &\hspace{1ex} $|1_A,1_B\rangle |\{0\}\rangle$
      & \hspace{1ex}$|1_A,0_B\rangle |1^{(2)}\rangle$
      & $D_{\mathrm {ix}}=(\omega_B+\omega')
        (\omega_A+\omega_B)
        (\omega_A+\omega)$\\
($8$)
      &  $|0_A,1_B\rangle |1^{(1)}\rangle$
      & \hspace{2ex}  $|1_A,1_B\rangle |\{0\}\rangle$
      & \hspace{1ex}$|0_A,1_B\rangle |1^{(2)}\rangle$
        & $D_{\mathrm {x}}=(\omega_B+\omega')
         (\omega_A+\omega_B)
            (\omega_B+\omega)$\\
($9$)
      &  $|0_A,1_B\rangle |1^{(1)}\rangle$
      & \hspace{2ex}$|1_A,1_B\rangle
        |1^{(2)},1^{(3)}\rangle$
      & \hspace{1ex}$|1_A,0_B\rangle |1^{(4)}\rangle$
      & $D_{\mathrm {xi}}=(\omega_B+\omega')
         (\omega_A+\omega_B+\omega'+\omega)
     (\omega_A+\omega')$\\
($10$)
      &  $|0_A,1_B\rangle |1^{(1)}\rangle$
      & \hspace{2ex} $|1_A,1_B\rangle |1^{(2)}, 1^{(3)}\rangle$
      & \hspace{1ex}$|0_A,1_B\rangle |1^{(4)}\rangle$
      & $D_{\mathrm {xii}}=(\omega_B+\omega')
         (\omega_A+\omega_B+\omega'+
         \omega)
     (\omega_B+\omega)$\\
\hline
\end{tabular}
\caption{
\label{Tab1}
Intermediate states contributing to the two-objects potential and
corresponding denominators.
}
\end{center}
\end{table}

%%%%%%%%%%%%%%%%%%%%%%%%%%%%%%%%%%

\end{document}